# On the (hydrodynamic) memory in the theory of Brownian motion

V. Lisy[1] and J. Tothova

Institute of Physics, P.J. Safarik University, Jesenna 5, 041 54 Kosice, Slovakia

E-mail: lisy@upjs.sk, tothova@upjs.sk

**Abstract.** The aim of this paper is to remember and review several exceptional investigations on the theory of Brownian motion. Although in these works the first correct hydrodynamic theories of the translational and rotational Brownian motion have been created, they remained unknown or very little known to the physical community and long time after the appearance of the original papers their main results were rediscovered by other authors. The reviewed works are highly interesting not only historically but from the methodical point of view as well.

---

[1] Author to whom any correspondence should be addressed.

# Contents

## Introduction

Brownian motion in suspensions has been the subject of investigations for nearly one hundred years. According to the western literature the modern period of these studies begins at the late sixtieth and early seventieth of the last century when the famous long-time "tails" of the molecular velocity autocorrelation function (VAF) (persistent or long-lived correlations) have been discovered. First in the computer experiments [1 - 3], later they have been confirmed theoretically (see e.g. [4 – 6] and the review [7]), and experimentally [8 – 13]. In particular, this discovery doubted the commonly accepted conception on the microscopic and macroscopic properties of liquids as characterized by very different time scales and has shown the real limits of the hydrodynamic theory. The common view on the history of this subject can be found in a number of papers, e.g. [14]. This view however almost completely forgets several substantial contributions to the theory. In some cases the principal results in the theory of the Brownian motion were rediscovered later and ascribed to other authors. First of all it concerns the considerable paper by Vladimirsky and Terletzky [15] to whom the first hydrodynamic theory of the Brownian motion belongs. Their theory will be presented in the next section. Later Leontovich [16] derived the so called generalized Langevin equation for the Brownian motion as a simple consequence of the fluctuation-dissipation theorem and obtained the correct expression for the mean square displacement (MSD) of the Brownian particle. Gitterman and Gertsenshtein [17] have generalized the results [15] to the case of Brownian particles in compressible environment. The mentioned computer experiments were explained by Fisher in 1970 [18] but the solution of an interesting phenomenon of the oscillations and decay of the molecular VAF observed by Rahman [1] was proposed already in 1967 [19]. The computer simulations have shown the existence of the long-time asymptotes (tails) $\sim t^{-3/2}$ in the VAF of a molecule in classical fluid. As is well known, the existence of these tails had far-reaching consequences on the physics of fluids and stimulated an intense flow of theoretical and experimental investigations of the collective correlations in liquids. The long-time asymptotes of the correlation functions describing the translational motion of Brownian particles have similar properties as the correlation functions for individual molecules. The effects of hydrodynamic memory appeared to be important also for the rotational Brownian motion. For the first time the correlation function of the angular velocity of the spherical Brownian particle has been derived by Zatovsky [20] who showed that the memory reveals in the long-time algebraic asymptote ~ $t^{-5/2}$ independent on the size and density of the particle. Analogous results have been obtained later, e.g. in Refs. [21 - 23], and confirmed by numerical simulations [24]. Long-time rotational motion of a rigid body of an arbitrary shape has been considered by Cichocki and Felderhof [25], the VAF of interacting Brownian particles was studied in Refs. [26], etc. - the rotational Brownian motion is the subject of continuous interest [29 - 31].

The hydrodynamic approach to the Brownian motion has essentially enriched the classical Einstein theory valid for the times $t \to \infty$ and showed the limitations of the later attempts to generalize it for arbitrary times (described in many textbooks and encyclopedias, see e.g. the popular Weisstein encyclopedia on the web [32] and below). This article is devoted to the presentation of two pioneering works, one dealing with the hydrodynamic theory of translational, the other with the rotational Brownian motion. Our aim was to remember these unknown or very little known works and demonstrate that correct hydrodynamic theories of the Brownian motion have been created long time before the equivalent results by other authors, which are traditionally cited in the literature. In our opinion, the reviewed works are highly interesting not only historically but from the methodical point of view as well.

## 1. Hydrodynamic theory of the translational Brownian motion: The forgotten work by Vladimirsky and Terletzky

As already mentioned, the first hydrodynamic theory of the Brownian motion has been built by Vladimirsky and Terletzky [15]. This work published in 1945 in the renowned Soviet Journal of Experimental and Theoretical Physics is almost completely neglected by the western authors. At the same time, in this paper the correct exact analytical solution for the MSD of the Brownian particle in incompressible fluid has been found. For the first time also the so called hydrodynamic (Ehrenfest) paradox is explained (the value of the VAF for the particle at the time $t = 0$ contradicts to the equipartition theorem). The priority of these results is ascribed to other authors [4, 6]. (In a number of papers one finds generalizations of the work [15] to compressible solvents [33, 34, 14]: also in this case the first paper where the hydrodynamic theory of the Brownian motion has been correctly generalized to the motion in an compressible fluid is "forgotten"; this is the work by Gitterman and Gertsenshtein [17].) Vladimirsky and Terletzky come from the Boussinesq equation for the friction force on a sphere during its motion in an incompressible fluid. Traditionally this resistance force is modeled by the Stokes force that, at a given time, is determined by the velocity of the body at the same moment of time. In general it is not correct: the Stokes force can be used only for the steady flow, i.e. at $t \to \infty$. Within the usual (linearized) nonstationary Navier-Stokes hydrodynamics this force should be replaced by the expression [35]

$$\vec{F}(t) = -6\pi\eta R \left\{ \vec{u}(t) + \frac{\rho R^2}{9\eta} \frac{d\vec{u}}{dt} + \sqrt{\frac{\rho}{\pi\eta}} \int_{-\infty}^{t} \frac{d\vec{u}}{d\tau} \frac{d\tau}{\sqrt{t-\tau}} \right\}, \qquad (1)$$

where $R$ is the particle radius, and $\rho$ and $\eta$ are the density and viscosity of the solvent. Equation (1) is valid for all times, except very short times when the compressibility effects have to be taken into account, i.e. $t \gg R/c$, where $c$ is the sound velocity is required. This expression has been derived by Boussinesq in 1885 [36] (also in [37]). In the literature (from the recent papers see e.g. [37 – 40] this force is often called the Basset history force [41] or the Basset-Oseen force [42], although the most appropriate name seems to be the Boussinesq-Basset force (for a discussion of this question see [43 – 45]). As it should be in the nonstationary case, the Boussinesq-Basset force on the particle at the time *t* is determined by its velocities and accelerations in all the preceding moments of time. This phenomenon can be called the viscous aftereffect or the hydrodynamic memory. It is seen from Eq. (1) that for fluids with the density comparable to the density of the Brownian particle (which is the usual case of neutrally buoyant particles) the additional terms (to the Stokes one) cannot be neglected since in the equation of motion for the particle they are of the same order as the inertial term.

The equation of motion for the Brownian particle is in the work [15] solved by the following elegant way. The authors use the theorem by Vladimirsky [46], which is a direct consequence of the general fluctuation-dissipation theorem [47]. According to the Vladimirsky's theorem the MSD of the Brownian particle $\langle \Delta x^2(t) \rangle$ is given by the equation of motion

$$M \frac{d^2}{dt^2} \langle \Delta x^2(t) \rangle - F(t) = 2k_B T, \tag{2}$$

where $M$ is the mass of the particle, $x$ is any of the Cartesian components of its position vector, and $F$ is the $x$ component of the Boussinesq force (1). The constant force at the right ($k_B$ is the Boltzmann constant and $T$ the temperature) begins to act on the particle at the time $t = 0$; up to this moment the particle is at rest together with the liquid. Equation (2) can be rewritten in the form

$$\left(M + \frac{M_s}{2}\right) \dot{V}(t) + 6\sqrt{\frac{M_s \xi}{2\pi}} \int_0^{\sqrt{t}} \dot{V}(t - \beta^2) d\beta + \xi V(t) = 2k_B T \tag{3}$$

with $M_s$ being the mass of the solvent displaced by the particle,

$$V(t) = \frac{d}{dt} \langle \Delta x^2(t) \rangle, \tag{4}$$

and $\xi = 6\pi\eta R$ is the Stokes friction coefficient. Let us introduce the characteristic times

$$\tau_R = \frac{R^2 \rho}{\eta}, \qquad \tau = \frac{M + M_s/2}{\xi} = \left(1 + 2\frac{M}{M_s}\right) \frac{\tau_R}{9}. \tag{5}$$

Equation (3) then takes the form

$$\dot{V}(t) + \frac{2}{\tau}\sqrt{\frac{\tau_R}{\pi}} \int_0^{\sqrt{t}} \dot{V}(t - \beta^2) d\beta + \frac{1}{\tau} V(t) = 2\Phi_0, \qquad \Phi_0 = \frac{k_B T}{M + M_s/2}. \tag{6}$$

The problem has to be solved with the evident initial conditions

$$V(0) = 0, \quad \langle \Delta x^2(0) \rangle = 0. \tag{7}$$

It is also seen from Eq. (6) that $\dot{V}(0) = 2\Phi_0$.

This formulation exactly corresponds to the problem of the motion of a sphere in a constant (gravitational) field that has been solved already by Boggio [48]. He used the Abel's integral relations (Abel's inversion) for the functions $\psi$ and $\varphi$, that can be rewritten here as follows:

$$\psi(t) = \frac{2}{\sqrt{\pi}} \int_0^{\sqrt{t}} \dot{\varphi}(t-\beta^2) d\beta, \quad \varphi(t) - \varphi(0) = \frac{2}{\sqrt{\pi}} \int_0^{\sqrt{t}} \psi(t-\beta^2) d\beta. \tag{8}$$

The function $\psi(t)$ can be chosen as

$$\psi(t) = \frac{2}{\sqrt{\pi}} \int_0^{\sqrt{t}} \dot{V}(t-\beta^2) d\beta = \frac{\tau}{\sqrt{\tau_R}} \left( 2\Phi_0 - \frac{1}{\tau} V(t) - \dot{V}(t) \right), \tag{9}$$

where the second equality follows from Eq. (6). From the Abel's relations and the initial condition for $V(t)$ one obtains

$$V(t) = \frac{2\tau}{\sqrt{\pi \tau_R}} \int_0^{\sqrt{t}} d\beta \left[ 2\Phi_0 - \frac{1}{\tau} V(t-\beta^2) - \dot{V}(t-\beta^2) \right]. \tag{10}$$

This expression has to be inserted in Eq. (6), derived with respect to the time and then the equation (9) for the function $\psi(t)$ is again used. This results in the ordinary differential equation for the function $V(t)$:

$$\ddot{V}(t) + \left( \frac{2}{\tau} - \frac{\tau_R}{\tau^2} \right) \dot{V}(t) + \frac{1}{\tau^2} V(t) = \frac{2\Phi_0}{\tau} \left( 1 - \sqrt{\frac{\tau_R}{\pi t}} \right). \tag{11}$$

This equation is solved by standard methods. The solution is searched for in the form $\sim \exp(\lambda^2 t)$, so that the characteristic equation for the (complex) quantities $\lambda$ is

$$\left( \lambda^2 + \frac{\sqrt{\tau_R}}{\tau} \lambda + \frac{1}{\tau} \right) \left( \lambda^2 - \frac{\sqrt{\tau_R}}{\tau} \lambda + \frac{1}{\tau} \right) = 0. \tag{12}$$

Two of the roots $\lambda_{1,2}$ of the quadratic expressions on the left are equal. It is thus sufficient to solve one of the equations, e.g. the first one. It is suitable to write the solution using the relations

$$\lambda_1 + \lambda_2 = \frac{\sqrt{\tau_R}}{\tau}, \quad \lambda_1 \lambda_2 = \frac{1}{\tau}. \tag{13}$$

The following relations are also valid:

$$\lambda_1^2 - \lambda_2^2 = -\frac{\tau_R}{\tau^2} \sqrt{1 - 4\frac{\tau}{\tau_R}}, \quad \lambda_1^2 + \lambda_2^2 = -\frac{\tau_R}{\tau^2} \left( 1 - 2\frac{\tau}{\tau_R} \right). \tag{14}$$

Equation (14) will be used in constructing the solution of Eq. (11) as a sum of the general solution of the homogeneous equation and a particular solution of the inhomogeneous equation. Using the initial conditions (7) we have

$$2\Phi_0 V(t) = \frac{1}{\lambda_1^2 \tau \lambda_2^2} + \frac{1}{\lambda_1 - \lambda_2} \left( \frac{1}{\lambda_1} e^{\lambda_1^2 t} - \frac{1}{\lambda_2} e^{\lambda_2^2 t} \right)$$

$$+\frac{1}{\sqrt{\pi}}\frac{1}{\lambda_1-\lambda_2}\left(e^{\lambda_2^2 t}\int_0^t \frac{dx}{\sqrt{x}}e^{-\lambda_2^2 x} - e^{\lambda_1^2 t}\int_0^t \frac{dx}{\sqrt{x}}e^{-\lambda_1^2 x}\right). \qquad (15)$$

The quantity $V(t)$ directly determines the time-dependent diffusion coefficient of the particle,

$$D(t) = \frac{1}{2}\frac{d}{dt}\langle \Delta x^2(t)\rangle = \frac{1}{2}V(t). \qquad (16)$$

The VAF is given by the relation

$$\Phi(t) = \langle \dot{x}(t)\dot{x}(0)\rangle = \frac{1}{2}\frac{d^2}{dt^2}\langle \Delta x^2(t)\rangle = \frac{1}{2}\dot{V}(t). \qquad (17)$$

Using the complementary error function $\mathrm{erfc}(z) = 1 - \mathrm{erf}(z)$ [49],

$$\mathrm{erf}(z) = \frac{2}{\sqrt{\pi}}\int_0^z e^{-x^2}dx,$$

one can express the function $\Phi(t)$ in the compact form due to Hinch [6]

$$\Phi(t) = \frac{\Phi_0}{\lambda_1-\lambda_2}\left\{\lambda_1 e^{\lambda_1^2 t}\mathrm{erfc}(\lambda_1\sqrt{t}) - \lambda_2 e^{\lambda_2^2 t}\mathrm{erfc}(\lambda_2\sqrt{t})\right\}, \quad \Phi_0 = \Phi(0), \qquad (18)$$

that is valid for an arbitrary value of the discriminant $D = \tau_R \tau^{-4}(\tau_R - 4\tau)$ of the characteristic equation (12). In the special case of equal roots when $\tau_R = 4\tau$,

$$\Phi(t) = \Phi_0\left\{-\frac{2}{\sqrt{\pi}}\sqrt{\frac{t}{\tau}} + \left(1+2\sqrt{\frac{t}{\tau}}\right)\exp\left(\frac{t}{\tau}\right)\mathrm{erfc}\left(\sqrt{\frac{t}{\tau}}\right)\right\}, \qquad (19)$$

and when the discriminant is negative, the solution is expressed through the error function of imaginary argument. Using its asymptotic expansion for large $|z|$ [49],

$$\sqrt{\pi}ze^{z^2}\mathrm{erfc}(z) \sim 1 + \sum_{m=1}^{\infty}(-1)^m \frac{(2m-1)!!}{(2z^2)^m},$$

we get for $\Phi(t)$, $t \to \infty$, the following formula:

$$\Phi(t) \sim \frac{\Phi_0}{2\sqrt{\pi}}\frac{\tau\sqrt{\tau_R}}{t^{3/2}}\left\{1 - \frac{3}{2}\left(1-2\frac{\tau}{\tau_R}\right)\frac{\tau_R}{t} + ...\right\}, \quad \Phi_0 \tau = D, \qquad (20)$$

where $D$ is the diffusion coefficient of the particle. The second term in the brackets, expressed through the mass of the particle and the displaced solvent, has the form $[7 - 4M/M_s]/(6t)$. Finally, the mean square displacement of the Brownian particle is found integrating the function $V(t)$ from 0 to $t$,

$$\langle \Delta x^2(t)\rangle = 2D\left\{t - 2\left(\frac{\tau_R t}{\pi}\right)^{1/2} + \tau_R - \tau + \frac{1}{\tau}\frac{1}{\lambda_1-\lambda_2}\left[\frac{1}{\lambda_1^3}e^{\lambda_1^2 t}\mathrm{erfc}(\lambda_1\sqrt{t}) - \frac{1}{\lambda_2^3}e^{\lambda_2^2 t}\mathrm{erfc}(\lambda_2\sqrt{t})\right]\right\} \qquad (21)$$

(it contains the time-independent term $\tau_R - \tau = 2\tau_R(4 - M/M_s)/9$). The asymptotic expansion of this equation for $t \to \infty$ is

$$\langle \Delta x^2(t) \rangle = 2Dt\left[1 - \frac{2}{\sqrt{\pi}}\left(\frac{\tau_R}{t}\right)^{1/2} + \frac{2}{9}\left(4 - \frac{M}{M_s}\right)\frac{\tau_R}{t} - \frac{1}{9\sqrt{\pi}}\left(7 - 4\frac{M}{M_s}\right)\left(\frac{\tau_R}{t}\right)^{3/2} + ...\right]. \quad (22)$$

In the literature the problem of the translational Brownian motion is solved by various different ways [6, 14, 33, 34]. Here we have demonstrated that the equations (18) – (22), which are usually used for the description of the Brownian motion in incompressible fluid, are fully equivalent to the original result [15] represented by Eq. (15).

It is discussed in Ref. [15] that the Ornstein result [50] presented in many textbooks,

$$\langle \Delta x^2(t) \rangle = \frac{2D}{M}\left(t - M\frac{1 - \exp(-t\xi/M)}{\xi}\right), \quad (23)$$

follows from the theory [15] only when the two conditions are simultaneously fulfilled: $M_s/2 \ll M$ and $t \ll \sqrt{MM_s/2}/\xi$, or at very long times $t \gg M/\xi$, when, however, the Einstein theory is valid. In the time scale between these limits the difference from these theories is more significant for large values $\tau_R$, i.e. for large particles and large density and small viscosity of the solvent.

## 2. On the first hydrodynamic theory of rotational Brownian motion

The next paper that we would like to shortly re-tell is the work by Zatovsky [20] dated 1969 (i.e. three years before the often cited work by Berne [22], see e.g. [25], whose result for the VAF of the Brownian particle was however not correct). In the Zatovsky's work, unknown to the western physicists, the first hydrodynamic theory of the rotational Brownian motion was built. That is, the correlation function of the angular velocity of the spherical Brownian particle in a liquid has been found taking into account the viscous aftereffect.

Usually the force acting on a rigid sphere of radius $R$ from the molecules of its environment is divided into the mean force - the friction force with the torque

$$\vec{M} = -8\pi R^3 \eta \vec{\Omega}, \quad (24)$$

where $\Omega$ is the angular velocity of the particle rotation, and the force of random hits with the zero mean value. The expression (24) is valid only for the steady motion. In the general case, the torque on the particle depends not only on the angular velocity of the particle at the given time, but also on the whole history of its motion. The rotational Brownian motion of a particle suspended in a liquid becomes a non-Markovian random process. The corresponding generalization is of self-dependent interest in the theory of the Brownian motion. It could also have implications on the interpretation of various phenomena in liquids since the analysis of the spectra of nuclear magnetic and electron resonance, Rayleigh and Raman scattering of light, and others, are often made using correlation functions of random rotational motion of molecules.

In Ref. [20] the correlation function of the angular velocity of the particle is considered,

$$\varphi(t) = \langle \vec{\Omega}(0)\vec{\Omega}(t) \rangle. \quad (25)$$

As usually, the angular brackets designate the averaging over the equilibrium ensemble of particles, related to the moment $t = 0$. According to the fluctuation-dissipation theorem the Fourier transform of the correlation function (25),

$$\tilde{\varphi}(\omega) = \frac{1}{\pi} \int_0^\infty dt\, \varphi(t) \cos \omega t \qquad (26)$$

is proportional to the real part of the complex "rotational mobility" $b(\omega)$ defined by the relation

$$\vec{\Omega}(\omega) = b(\omega) \vec{M}^b(\omega), \qquad (27)$$

so that

$$\tilde{\varphi}(\omega) = \frac{3k_B T}{\pi} \operatorname{Re} b(\omega). \qquad (28)$$

The problem is thus reduced to the search for the dependence of the mobility on the frequency $\omega$. Again, for not too short times the liquid can be considered as incompressible. The Fourier transform of the torque of friction forces on the sphere is [51]

$$\vec{M}(\omega) = -8\pi R^3 \eta \left[ 1 - \frac{1}{3} \frac{i\omega \tau_R}{1 + \sqrt{-i\omega \tau_R}} \right] \vec{\Omega}(\omega), \qquad (29)$$

where $\tau_R = R^2 \rho / \eta$ and $\sqrt{-i} = (1-i)/\sqrt{2}$. When also the external moment $M^b$ (Eq. 27) acts on the sphere, the equation of motion for the mean angular velocity takes the form

$$-i\omega \vec{\Omega}(\omega) I + 8\pi R^3 \eta \left[ 1 - \frac{1}{3} \frac{i\omega \tau_R}{1 + \sqrt{-i\omega \tau_R}} \right] \vec{\Omega}(\omega) = \vec{M}^b(\omega), \qquad (30)$$

where $I$ is the moment of inertia of the sphere. From here it has been found for the rotational mobility

$$b(\omega) = \frac{\tau_R}{I} \frac{1 + \sqrt{-i\omega \tau_R}}{3x + 3x\sqrt{-i\omega \tau_R} - i\omega \tau_R (1+x) - i\omega \tau_R \sqrt{-i\omega \tau_R}}. \qquad (31)$$

Here $x = 5\rho/\rho_1$, and $\rho_1$ is density of the particle.

The searched correlation function is determined by the inverse transform of Eq. (28). Taking into account (31) it equals to

$$\varphi(t) = \frac{3k_B T}{\pi I} \int_{-\infty}^{\infty} dy \cos\left( y \frac{t}{\tau_R} \right) \frac{1 + \sqrt{iy}}{3x + 3x\sqrt{iy} + (1+x)iy + iy\sqrt{iy}}. \qquad (32)$$

The result of integration depends on the behavior of the roots of the denominator in the integrand. The character of the roots of the cubic equation

$$z^3 + (1+x)z^2 + 3xz + 3x = 0 \qquad (33)$$

is determined by the discriminant $D$

$$D = p^3 + q^2, \qquad p = x - \left(\frac{1+x}{3}\right)^2, \qquad q = x - \frac{x^2}{2} + \left(\frac{1+x}{3}\right)^3. \tag{34}$$

When $D < 0$, Eq. (33) has three real roots depending only on the parameter $x$. If $D = 0$, it has three real roots, two of which are identical, and when $D > 0$, three different roots, one of which is real and the other two are complexly conjugated. From (34) one finds $D(x_0) = 0$ at $x_0 \approx 45$. This corresponds to $\rho_1 \approx \rho/9$. In Ref. [20] the case of not too light particles is considered in detail. In this case $\rho_1 > \rho/9$ and $D > 0$. Let $z_1$ and $z_2$, $z_2^*$ are the roots of Eq. (33). The fraction in the integral (32) can be then expressed as a sum of simplest fractions

$$A\left[\frac{1}{\sqrt{iy} - z_1} - \frac{1}{2}\left(\frac{1 - iB}{\sqrt{iy} - z_2} + \frac{1 + iB}{\sqrt{iy} - z_2^*}\right)\right], \tag{35}$$

$$A = \frac{1 + z_1}{|z_2|^2 + z_1^2 - z_1(z_2 + z_2^*)}, \qquad B = \frac{z_1}{z_2 - z_2^*}\left[\frac{1}{A} + z_1 - \frac{z_2 + z_2^*}{2}\right].$$

The integration (32) with the use of (35) leads to the following correlation function:

$$\varphi(t) = \frac{3k_B T}{I} A\left[z_1 w\left(-iz_1\sqrt{\frac{t}{\tau_R}}\right) - \mathrm{Re}(1 - iB) z_2 w\left(-iz_2\sqrt{\frac{t}{\tau_R}}\right)\right], \tag{36}$$

where the function $w(z)$ for the complex argument [49]

$$w(z) = e^{-z^2}\left(1 + \frac{2i}{\sqrt{\pi}}\int_0^z e^{t^2} dt\right)$$

has been studied in detail and tabulated in [52].

At $t = 0$ it follows from (25) and (36) for the mean square angular velocity

$$\langle \Omega^2 \rangle = \frac{3k_B T}{I}. \tag{37}$$

The correlation time of the angular velocity can be defined as

$$\tau_c = \int_0^\infty \frac{\varphi(t)}{\varphi(0)} dt = \frac{\tau_R}{3x}. \tag{38}$$

Using the expansion of the complex error function for small $|z|$ and its asymptotic representation for large $|z|$ [49, 52], the following expressions have been found:

$$\varphi(t) = \frac{3k_B T}{I}\left\{1 - 2x\sqrt{\frac{t}{\pi\tau_R}} + (3 - x - x^2)\frac{t}{\tau_R} + \ldots\right\}, \qquad \frac{t}{\tau_R} \ll 1, \tag{39}$$

$$\varphi(t) = \frac{3k_B T}{32\pi\sqrt{\pi\rho}} \left(\frac{\rho}{\eta t}\right)^{5/2} \left\{ 1 - \frac{45xA}{2} \left( \text{Re} \frac{1-iB}{z_2^6} - \frac{1}{z_1^6} \right) \frac{\tau_R}{t} + \ldots \right\}, \quad \frac{t}{\tau_R} \gg 1. \tag{40}$$

In the usual theory built on the expression (24) the correlation function for the angular velocity has the Debye form [53]

$$\varphi_D(t) = \frac{3k_B T}{I} \exp\left(-\frac{t}{\tau_c}\right). \tag{41}$$

Equations (36), (39) and (40) significantly differ from (41). Only in the case of very heavy particles when $\sqrt{\rho/\rho_1} \ll 1$, and for $t \gg \tau_R$ simple estimations of the roots from Eq. (33) give

$$z_1 \approx -1, \quad z_2(z_2^*) \approx \pm i\sqrt{3x},$$

so that the solution (36) becomes

$$\varphi(t) = \frac{3k_B T}{I} \exp\left(-3x \frac{t}{\tau_R}\right) + O(\sqrt{x}),$$

which corresponds to (41).

It was noted in [20] that the main term of the asymptotic representation of the correlation function does not depend on the size of the particles and on their density. In connection with this the author proposes another derivation of this term. The vibrational rotation of a particle in a viscous liquid at small angular velocities, due to the stick boundary conditions at its surface, occurs likewise the motion of a small portion of the surrounding liquid. The time behavior of the angular velocity, and consequently the correlation function at long times should asymptotically correspond to the behavior of the translational velocity vortex of the motion of liquid in the same point. Solving the Navier-Stokes equations, it has been shown that the correlation function of the vortex, $\Phi(t-t'; |\vec{r}-\vec{r}'|) = (1/4)\langle \text{rot}\,\vec{v}(\vec{r},t)\,\text{rot}\,\vec{v}(\vec{r}',t')\rangle$, is at $\vec{r} = \vec{r}'$ identical with the main term in Eq. (40), i.e.

$$\Phi(t,0) = \frac{3k_B T}{32\pi\sqrt{\pi\rho}} \left(\frac{\rho}{\eta t}\right)^{5/2}. \tag{42}$$

Finally, the author illustrates the essential difference between the previous theory and the theory that takes into account the viscous aftereffect numerically, on the special case of the rotation of a large particle among the particles with the same density. It is concluded that the account for the viscous aftereffect significantly changes the results of the theory of rotational Brownian motion of a particle suspended in a liquid.

## 3. Conclusion

The theory of Brownian motion is of extraordinary importance for various fields of physics, as well as for chemical, biological, and other sciences. After the creation of the first theory of the Brownian motion by Einstein in 1905 it has been significantly developed and is still developing. Nowadays it can be considered as a self-existing part of the science. Particular aspects of the theory have been many times discussed and reviewed in the literature. This little review does not summarize new results achieved in recent years. As opposite, we have rewritten here for the readers two old papers, one of which [15] appeared sixty and the other [20] thirty five years ago. We decided to remember these papers for several reasons. The main reason is that in spite of the fact that in these remarkable works the first and correct hydrodynamic theories of the translational and rotational Brownian motion have been built, they were almost totally ignored. Instead, equivalent results obtained by other authors are used and cited in the literature. Due to the importance of these results, the number of such citations is very high. Such unfairness is not a curiosity in the science and probably nothing will change after the appearance of this article. Nevertheless, we felt that we have to make an attempt to bring these papers to the eventual readers' attention. Not only because of their historical value but also because the methods of solution are elegant and despite the passed years still instructive. That is why we re-tell the papers in details. Several other little known papers in which significant results on the subject have been obtained are also mentioned.